\documentclass[twocolumn,aps,prx,showpacs,amsmath,amssymb,floatfix,longbibliography,superscriptaddress,eprint]{revtex4-1}

\usepackage{subfigure}

\usepackage{xcolor}
\definecolor{darkblue}{RGB}{0,0,150}
\definecolor{nightblue}{RGB}{0,0,100}

\usepackage{graphicx,mathtools,bm,bbm}
\usepackage{MnSymbol}
\usepackage[
colorlinks,
citecolor=darkblue,
linkcolor=darkblue,
urlcolor=nightblue]{hyperref}

\usepackage[english]{babel}
\usepackage[babel,kerning=true,spacing=true]{microtype}
\usepackage[utf8]{inputenc}

\usepackage{soul}

\newcommand{\vpd}{\vphantom{\dagger}}

\begin{document}



\title{Gauged cooling of topological excitations and emergent fermions on quantum simulators}
    
    
\author{Gilad Kishony}
\affiliation{Department of Condensed Matter Physics,
Weizmann Institute of Science,
Rehovot 76100, Israel}
\author{Mark S. Rudner}
\affiliation{Department of Physics, University of Washington, Seattle, WA 98195-1560, USA}
\author{Achim Rosch}
\affiliation{Institute for Theoretical Physics, University of Cologne, 50937 Cologne, Germany}
\author{Erez Berg}
\affiliation{Department of Condensed Matter Physics,
Weizmann Institute of Science,
Rehovot 76100, Israel}

\begin{abstract}
Simulated cooling is a robust method for preparing low-energy states of many-body Hamiltonians on near-term quantum simulators. In such schemes, a subset of the simulator's spins (or qubits) are treated as a ``bath,'' which extracts energy and entropy from the system of interest. However, such protocols are inefficient when applied to systems whose excitations are highly non-local in terms of the microscopic degrees of freedom, such as topological phases of matter; such excitations are difficult to extract by a local coupling to a bath. We explore a route to overcome this obstacle by encoding of the microscopic degrees of freedom into those of the quantum simulator in a non-local manner. To illustrate the approach, we show how to efficiently cool the ferromagnetic phase of the quantum Ising model, whose excitations are domain walls, via a 
``gauged cooling'' protocol in which the Ising spins are coupled to a $Z_2$ gauge field that simultaneously acts as a reservoir for removing excitations. 
We show that our protocol can prepare the ground states of the ferromagnetic and paramagnetic phases equally efficiently. 
The gauged cooling protocol naturally extends to (interacting) fermionic systems, where it is equivalent to cooling by coupling to a fermionic bath via single-fermion hopping. 
\end{abstract}

\maketitle


One of the principle near-term prospects of quantum computing is its application to quantum simulation \cite{2012CiracZoller, Georgescu_2014}. This umbrella term refers to the quantitative study of the quantum properties of microscopic particles and is directly applicable to the fields of materials science, high-energy physics, and quantum chemistry. Importantly,  quantum simulations might allow us to determine not only dynamics, but also ground state properties.

Several methods for preparation of ground states on quantum simulators have been proposed. 
These include variational methods \cite{McClean_2016, Moll_2018, Tilly_2022} which require 
a variational ansatz, 
and rely on classical optimization to minimize the energy. 
Others are adiabatic processes \cite{farhi2000quantum, Childs_2001, Aspuru_Guzik_2005, Albash_2018}, which 
are limited by convergence times that grow 
with system size when phase transitions are encountered.  
A third approach implements effective imaginary time evolution \cite{Motta_2019, Jouzdani_2022, Lin2021Compressed} 
by utilizing intermediate steps of tomography and costly classical post-processing.

\begin{figure}[t]
\begin{centering}
\includegraphics[width=0.75\columnwidth]{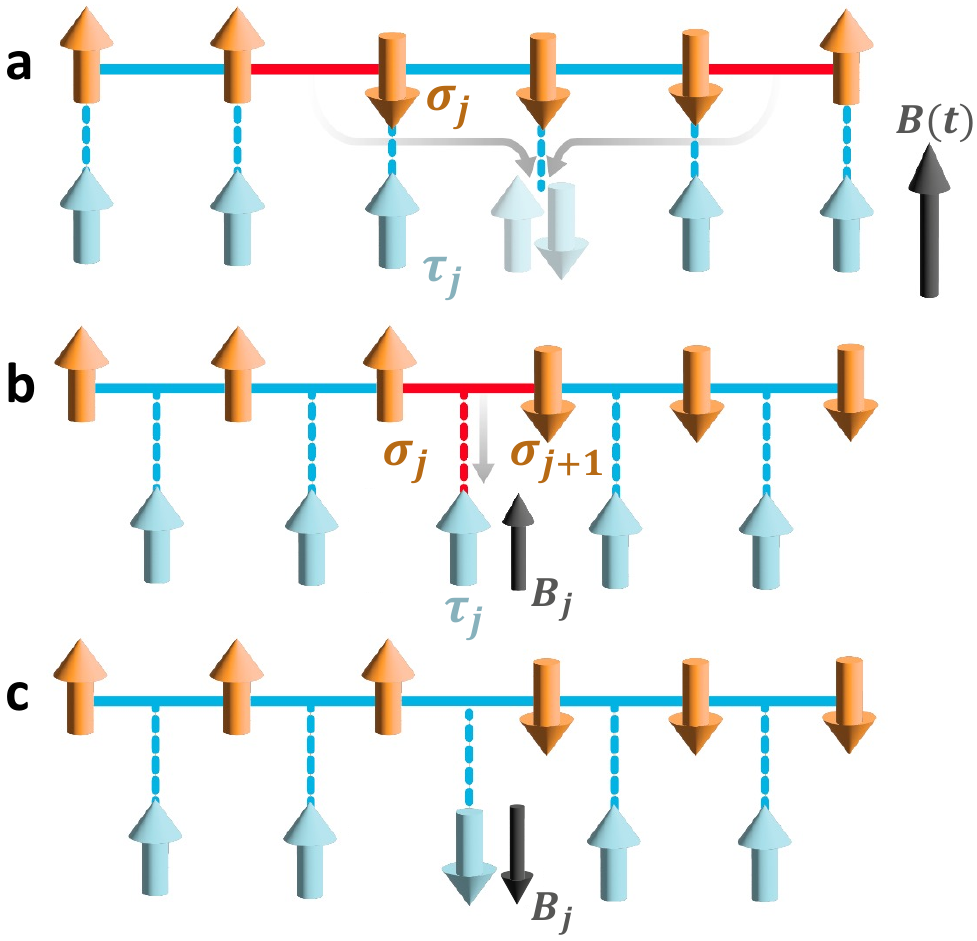}
\par\end{centering}
\caption{\textbf{(a)} Protocol for simulated adiabatic cooling by locally coupling the system spins $\sigma_j$ to bath spins $\tau_j$~\cite{mathhies2022adibatic_demag}, subject to a Zeeman field $B(t)$ which is gradually ramped down. The $\tau_j$'s are then reset to the fully polarized state, and a new cooling cycle begins. If the system is in a $Z_2$ symmetry-broken (ferromagnetic) phase, domain wall excitations (red) can only be removed in pairs. 
This dramatically slows down the cooling process. 
\textbf{(b)} A ``gauged cooling'' protocol where $\tau^z_j$ play the role of $Z_2$ gauge fields coupling to the bonds. 
The $\tau_j$'s are subject to a local Zeeman field $B_j$, whose sign varies from site to site, and whose magnitude decreases adiabatically. If a domain wall is present, $\tau^z_j$ on the same bond tends to flip to reduce the 
energy. \textbf{(c)} After $\tau^z_j$ has flipped, the domain wall is removed. The resulting state is equivalent to the ground state of the original 
model by a gauge transformation, which can be constructed after by measuring the $\tau^z_j$'s. Thus, a single domain wall has effectively been removed by a local operation. In the next cooling cycle, the sign of $B_j$ is chosen according to 
$\tau^z_j$, such that the $\tau_j$ spins are initially 
parallel to $B_j$.}
\label{fig:1}
\end{figure}

A fourth, recently proposed, class of algorithms (such as those in Refs. \cite{Boykin_2002, Kaplan2017, Metcalf_2020, Polla_2021, Zaletel_2021,  RodriguezThesis, mathhies2022adibatic_demag}) implement simulated cooling by
treating part of the degrees of freedom of the quantum simulator as an artificial ``bath,'' 
employing the idea of bath engineering~\cite{Kraus_2008, verstraete2008quantum, harrington2022engineered, mi2023stable}. 
In each cycle of the algorithm, the system is cooled by its coupling to the bath, whose state is later monitored and reset. While these protocols can be flexibly applied to any 
Hamiltonian, 
they are limited when it comes to cooling of topological excitations which are non-local in terms of the hardware's degrees of freedom 
\footnote{Throughout the discussion, we are assuming interactions are spatially local, or at least $k-$local with a finite $k$.}. Single topological excitations cannot be removed by local 
operations; several such excitations must coincide in order to be annihilated. 

The simplest example of such non-local excitations are domain walls in the ordered phase of the one-dimensional Ising model. Indeed, in the simulated adiabatic cooling protocol of Ref.~\cite{mathhies2022adibatic_demag}, it was found that cooling this phase is {\it parametrically} more difficult than cooling the paramagnetic phase, in the sense that the energy density of the system decreases as a power law with the number of cooling steps (as opposed to exponentially in the paramagnetic phase). 
This obstacle of can be overcome by breaking the $Z_2$ spin flip symmetry of the Ising model during the cooling process, which confines the domain wall excitations; however, such a solution is not available in other cases, such as topologically ordered phases, whose excitations are intrinsically non-local in nature.  

In this work, we propose a simulated cooling protocol that solves the problem of cooling non-local excitations, preparing the ferromagnetic and the paramagnetic phases of the quantum Ising model equally efficiently, by a non-local encoding of the physical degrees of freedom into those of the quantum simulator (Fig.~\ref{fig:1}). 
Rather than simulating the original Ising model, we simulate the gauged Ising model, where the Ising spins are coupled to a static $Z_2$ gauge field. This model is 
equivalent to the original quantum Ising model via a (non-local) unitary transformation~\footnote{Some care is required when handling systems with periodic boundary conditions, as we elaborate on below.}. 
During each cooling cycle, the gauge fields play the role of the bath, and are used to extract energy and entropy from the system. At the end of the cycle, the gauge fields are measured, and the measurement outcome is used to determine the Hamiltonian used in the next cooling cycle. 
We show that this protocol is highly efficient in both the paramagnetic and the ferromagnetic phases; in particular, in the ferromagnet, the gauge coupling allows to eliminate single domain walls locally. 

Interestingly, by mapping the system's spins into fermions, we show that our protocol is equivalent to cooling a many-body fermionic system by coupling it to a fermionic reservoir via single-fermion hopping. We analyze the performance of the protocol both in the solvable free-fermion limit, and away from this limit, and in the presence of decoherence. 

\begin{figure}[t]
\begin{centering}
\includegraphics[width=1.0\columnwidth]{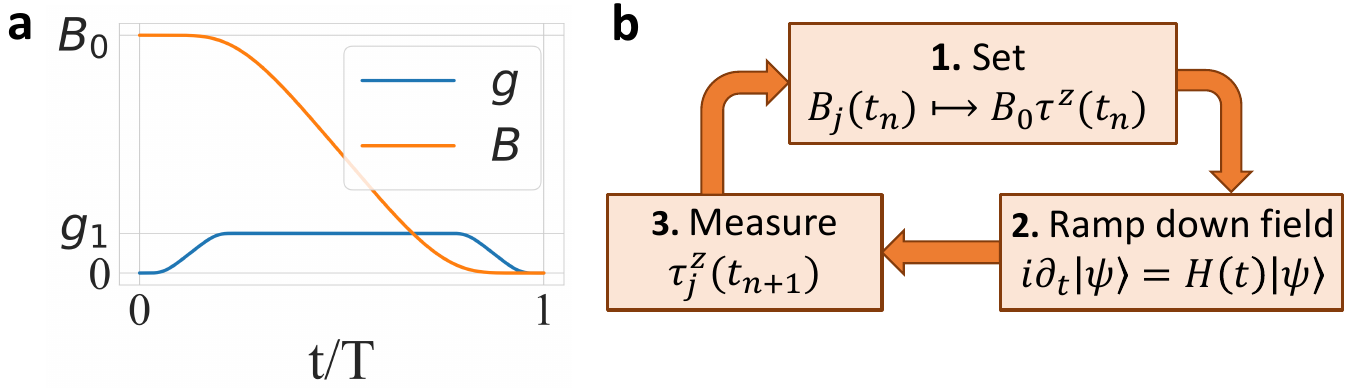}
\par\end{centering}
\caption{\textbf{(a)} The functions $B(t)$, $g(t)$ used in $H_{\rm c}$ [Eq.~(\ref{eq:Hc})]. 
The explicit functions appear in~\cite{SM}. 
\textbf{(b)} Gauged cooling protocol. The signs of the Zeeman fields $B_j$ are adjusted according to the result of the measurements of $\tau^z_j$ from the previous cycle. The system is then unitarily time evolved according to $H(t) = H_{\rm GTFI} + H_{\rm c}(t)$ [Eqs.~(\ref{eq:Hc}) and (\ref{eq:HG})] for a time $T$. The $\tau^z_j$'s are measured, and a new cycle begins. }
\label{fig:protocol}
\end{figure}

To set up the problem, we begin with a brief recap of the cooling protocol of Ref.~\cite{mathhies2022adibatic_demag}, applied to the quantum Ising model. Similar protocols were proposed in Refs.~\cite{Polla_2021, Zaletel_2021, Metcalf_2020, Boykin_2002, RodriguezThesis}. Each site of a one dimensional chain contains two spins, a ``system spin'' $\sigma_j$ and a ``bath spin'' $\tau_j$. The Hamiltonian is given by $H=H_{\rm TFIM}+H_{{\rm c}}$, where $H_{{\rm TFIM}}=-\sum_{j}\left[J\sigma_{j}^{x}\sigma_{j+1}^{x}+h\sigma_{j}^{z}\right]$ is the usual transverse field Ising model (TFIM) Hamiltonian, and the ``control Hamiltonian'' is given by 
\begin{equation}
    H_{{\rm c}}(t)=-\sum_{j}\left[B_j(t)\tau_{j}^{z}+g(t)\sigma_{j}^{x}\tau_{j}^{x}\right].
    \label{eq:Hc}
\end{equation}
In each cycle, the bath spins are initialized in the $\left\vert+1\right\rangle$ state in the $\tau^z_j$ basis. The system is time evolved by $H(t)$ for a period of time $T$, during which the ``Zeeman field'' $B_j(t) = B(t)$ applied to the bath spins 
is gradually ramped down while $g(t)>0$. Towards the end of the cycle, $g(t)$ is turned off. The profiles of $B(t)$ and $g(t)$ used in the protocol are shown in Fig. \ref{fig:protocol}(a). 
The bath spins are then reset back to $ \left\vert+1\right\rangle$, and a new cycle begins. In the large $T$ limit, such that the time evolution is adiabatic with respect to the gap of $H_{{\rm TFIM}}$, the steady state of the process (restricted to the $\sigma_j$ spins) approaches the ground state of $H_{{\rm TFIM}}$~\cite{mathhies2022adibatic_demag}. 
The process is directly analogous to adiabatic demagnetization of solid state systems, where the electronic and lattice degrees of freedom are cooled by coupling to electronic or nuclear spins, which are initially nearly fully polarized~\cite{Debye1926,Giauque1927}.

In the paramagnetic phase of the TFIM ($|h|>|J|$), the energy density of the state decreases exponentially with the number of cycles performed, $n_{\rm cyc}$; however, in the ferromagnetic phase ($|h|<|J|$), the energy density decreases as $1/n_{\rm cyc}$. The difference in performance in the two phases originates from the difficulty in cooling domain wall excitations by a local coupling to the bath [see Fig.~\ref{fig:1}(a)]. In the presence of a non-zero decoherence rate $\eta$, there is also a parametric difference in performance: in the paramagnet, the energy density in the steady state is proportional to $\eta$, whereas in the ferromagnet it is proportional to $\sqrt{\eta}$~\cite{mathhies2022adibatic_demag,SM}. 

To be able to remove both local (spin flip) and non-local (domain wall) excitations, we design an alternative protocol. We couple the $\sigma_j$'s to a static $Z_2$ gauge field, identified with $\tau^z_j$. The resulting gauged TFIM Hamiltonian,
\begin{equation}
    H_{{\rm GTFI}}=-\sum_{j}\left[J\sigma_{j}^{x}\tau_{j}^{z}\sigma_{j+1}^{x}+h\sigma_{j}^{z}\right],
    \label{eq:HG}
\end{equation}
commutes with $\{\tau_j^z\}$. In addition, $H_{{\rm GTFI}}$ is invariant under any local gauge transformation, generated by $G_j = \tau_{j-1}^x \sigma_j^z \tau_j^x$. For open boundary conditions, \eqref{eq:HG} is equivalent by a gauge transformation to $H_{\rm TFIM}$ in every sector of fixed $\tau^z_j$~\cite{Mcgreevyqft}. 
The transformation from a sector with a given $\{\tau^z_j\}$ to the sector 
$\{\tau^z_j=1\}$, where \eqref{eq:HG} is equivalent to $H_{\rm TFIM}$, is highly non-local; it is the non-local encoding of the degrees of freedom of $H_{\rm TFIM}$ that allows removing domain wall excitations by local interactions. 
Given the ability to cool to the ground state in a specific gauge configuration and to measure $\{\tau^z_j\}$, we can recover all ground state observables of the original TFIM problem. For a discussion of the effects of imperfect measurements, see \cite{SM}.

Importantly, the gauge degrees of freedom play dual roles in our protocol. They encode information about the original Ising spins, since the transformation from the TFIM to the gauged TFIM depends on $\{\tau^z_j\}$. In addition, they serve as the bath degrees of freedom used to cool the system, as we now discuss.

The 
cooling protocol consists of repeated cycles.  
Initially, the $\sigma$ spins 
are in an arbitrary (possibly unknown) state. 
Due to either an initialization step or a round of projective measurements, the initial state of the $\tau$ spins in each cycle is a known product state in the $\tau_{j}^{z}$ basis. 
The following steps
are performed in each cycle $n$, beginning at time $t_n = nT$: 

\vspace{0.25cm}
\begin{enumerate}
\item The ``Zeeman field'' $B_{j}$ is turned on to a value $B_0\tau_{j}^{z}(t_n)$,
where $B_{0}>0$ is chosen to be large compared to the size of a typical
term in the system's Hamiltonian ($|h|$ and $|J|$), and $\tau_j^z(t_n) = \pm 1$ refers to the eigenvalue of $\tau_j^z$ in the state at the beginning of cycle $n$. 
This choice of field orientations, along with $g(t_n) = 0$, ensures that each 
$\tau_{j}$ begins each cycle from its 
ground state, decoupled from the $\sigma$ spins. 
\item The 
coupling $g(t)$ is adiabatically ramped up to some finite
value and held fixed, while $B_{j}(t) = B(t)\tau_{j}^{z}(t_n)$ 
is adiabatically ramped to zero.
Here, adiabaticity is defined relative to the typical
energy scale of the system (e.g., the gap between the ground state
and the first excited state, if there is a non-zero gap in the thermodynamic
limit).
The coupling $g(t)$ is then adiabatically ramped
to zero.  Note that, while the $B$ and $g$ terms in Eq. \eqref{eq:Hc} break the gauge invariance of $H_{GTFI}$, gauge symmetry is restored at the end of the cycle when $B=g=0$.
\item 
The $\tau$ 
qubits are measured in the $\tau_{j}^{z}$ basis,
and another cycle begins.
\end{enumerate}
The profiles $B(t-t_n)$ and $g(t-t_n)$ are shown in Fig. \ref{fig:protocol}(a) and the steps of the protocol are illustrated in Fig. \ref{fig:protocol}(b).



\begin{figure}[t]
\begin{centering}
\includegraphics[width=0.8\columnwidth]{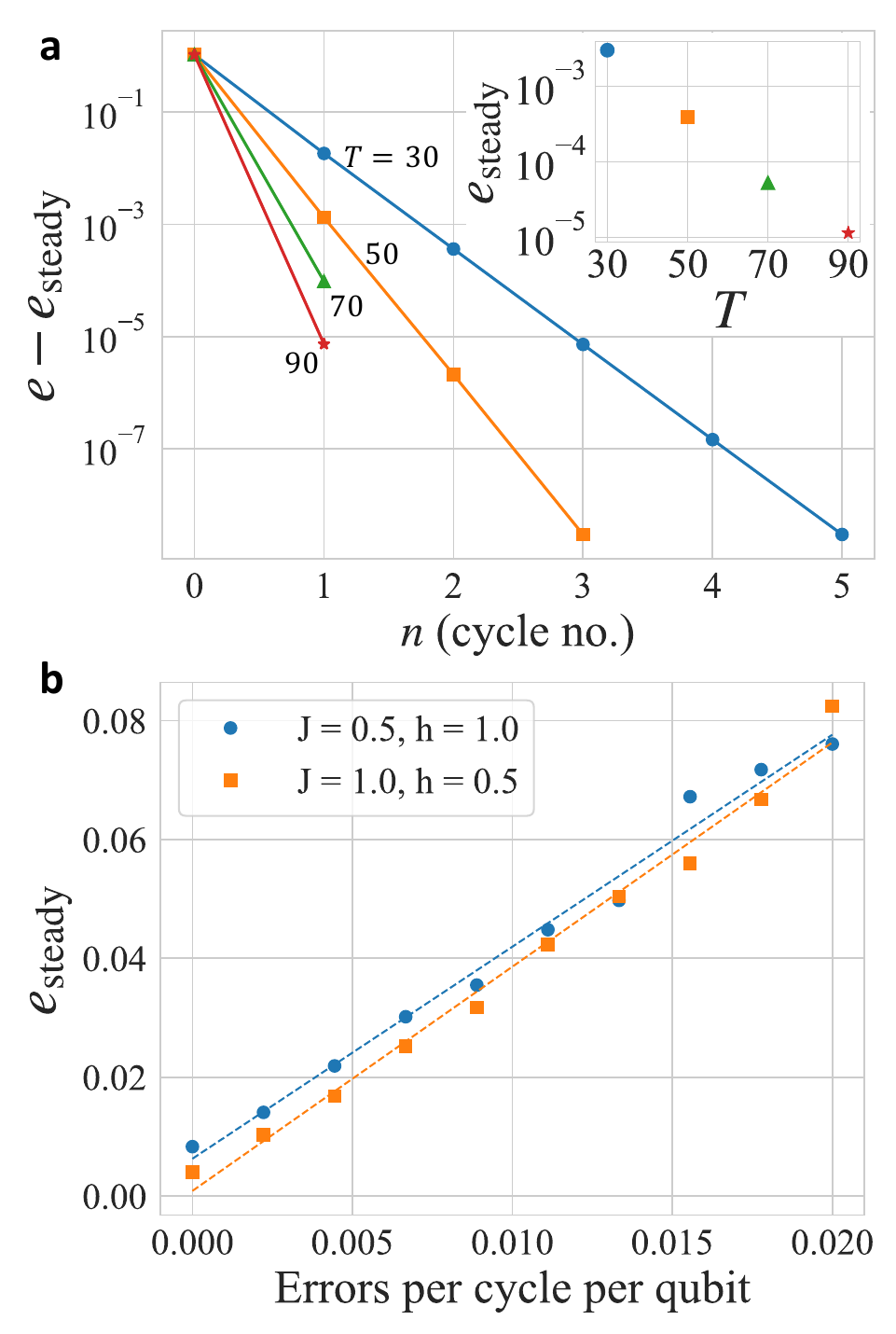}
\par\end{centering}
\caption{Numerical simulations in a system with $N=100$ sites and open boundary conditions performed using the free fermion representation (see \cite{SM}). The functions $B(t)$, $g(t)$ appear in \cite{SM}. The protocol parameters used were $g_1=0.5$, $B_0=3$. \textbf{(a)} Deviation of the energy density $e$ from the steady state value $e_{\rm{steady}}$ vs. protocol cycles in the ferromagnetic phase ($J=1$, $h=0.5$), shown for different values of $T$. Inset: convergence of the steady state energy density to the ground state value with increasing $T$. The initial state of the $\sigma$ spins is a maximally mixed state. The number of Trotter steps $N_t=10^5$ is sufficiently large such that the deviation of the steady state from the ground state is due to adiabatic errors. \textbf{(b)} Energy density of the steady state in the presence of depolarizing noise, in the paramagnetic (circles) and ferromagnetic (squares) phases, as a function of the mean number of errors per spin per cycle. Decoherence is applied to both the $\sigma$ and $\tau$ spins. Protocol parameters: $T=50$, $N_t=100$. The energies are averaged over $1000$ cycles per data point. In both phases, the energy density is approximately linear in the error rate.}
\label{fig:numerics}
\end{figure}

Remarkably, this protocol efficiently cools both the paramagnetic and the ferromagnetic phases, in the sense that in both phases the energy density decreases exponentially with $n_{\rm cyc}$, and in the presence of a decoherence rate $\eta$, the energy density in both phases is proportional to $\eta$. This is demonstrated by numerical simulations, as shown in Fig.~\ref{fig:numerics}. 

The reason the same efficiency is achieved in both 
phases is that the gauged cooling protocol can remove single domain walls of the ferromagnet as well as single spin flips in the paramagnet. In the ferromagnet, domain walls are removed simply by flipping a single gauge field $\tau^z_j$, effectively turning the corresponding bond from ferromagnetic to antiferromagnetic [see Fig.~\ref{fig:1}(b-c)]. The Hamiltonian with flipped $\tau^z_j$ is equivalent by a gauge transformation to the original one, but the domain wall has been eliminated. Note that this happens automatically by the tendency of the protocol to decrease the system's energy and entropy, realizing a form of autonomous error correction~\cite{Kapit2016,RodriguezThesis}. 

Importantly, unlike in the protocol of Ref.~\cite{mathhies2022adibatic_demag}, the $\tau_j$ spins are {\it not} reset at the end of the cycle. A reset operation does not commute with $H_{\rm GTFI}$, and would generically create new excitations. Rather, the sign of the field $B_j$ is adjusted in the next cycle according to the result of the measurement, such that the $\tau_j$ spins are initially in the ground state of the Zeeman part of $H_{\rm c}$.  Thus, the protocol presented here requires mid-circuit measurements and feedback. 
For platforms where mid-circuit measurements and feedback are prohibitive, it is possible to modify the protocol such that mid-circuit measurements are avoided, at the cost of introducing an additional ancilla qubit in each site, which is reset in every cycle (see \cite{SM}). 

An extension to a system with periodic boundary conditions is also possible. In this case, not all configurations of the $\tau_j^z$'s are equivalent up to a gauge transformation; the total $Z_2$ flux through the system, $F = \prod_j \tau_j^z$, is a gauge-invariant quantity. This quantity is known at the end of each cycle, since all the $\tau^z_j$ operators are measured. Cooling into the ground state of the quantum Ising model with periodic boundary conditions can be done by monitoring the flux sector $F$ after each cycle, and only performing measurements on the system in cycles where $F$ is $1$. In the ferromagnetic phase, this sector has a lower energy, so the system will naturally tend to settle into the $F=1$ sector (see \cite{SM}). Decoherence can flip the system between the two sectors, and then the measurements at the end of the cycle can be used to determine the global $Z_2$ flux.    

It it instructive to cast the cooling protocol in a fermionic language, by using a Jordan-Wigner transformation to represent the spins $\sigma_j$ and $\tau_j$ in terms of complex fermions, $a_j$ and $b_j$, as follows:
\begin{align}    
\label{eq:JW}
\sigma_{j}^{z}&=2a_{j}^{\dagger}a^{\vpd}_{j}-1,\,\tau_{j}^{z}=2b_{j}^{\dagger}b^{\vpd}_{j}-1,\\\sigma_{j}^{-}&=e^{i\pi\sum_{l<j}\left(a_{l}^{\dagger}a^{\vpd}_{l}+b_{l}^{\dagger}b^{\vpd}_{l}\right)}a_{j},\,\tau_{j}^{-}=e^{i\pi\sum_{l\le j}\left(a_{l}^{\dagger}a^{\vpd}_{l}+b_{l-1}^{\dagger}b^{\vpd}_{l-1}\right)}b_{j}.\nonumber
\end{align}
In terms of $a_j$ and $b_j$, the Hamiltonians $H_{\rm GTFI}$,  $H_{\rm c}(t)$ become
\begin{align}    
    &H_{\rm GTFI} = \sum_{j}\left[J\left(a_{j}^{\dagger}a^{\vphantom{\dagger}}_{j+1}+a_{j}^{\dagger}a_{j+1}^{\dagger}\right)-ha_{j}^{\dagger}a^{\vphantom{\dagger}}_{j}\right]+{\rm h.c.},\nonumber\\
    &H_{\rm c} = -\sum_{j}\left[B_{j}(t)b_{j}^{\dagger}b^{\vphantom{\dagger}}_{j}+g(t)\left(a_{j}^{\dagger}b^{\vphantom{\dagger}}_{j}+a_{j}^{\dagger}b_{j}^{\dagger}\right)\right]+{\rm h.c.}
    \label{eq:HG fermionic}
\end{align}
In terms of the $a_j$ fermions, the Hamiltonian $H_{\rm GTFI}$ describes a one-dimensional spinless mean-field superconductor; the ferromagnet ($h<J$) and paramagnet ($h>J$) correspond to the topological and the trivial phases, where the former phase supports Majorana edge modes, while the latter phase does not~\cite{kitaev_unpaired_2001}.
The control Hamiltonian $H_{\rm c}$ contains two terms: an on-site energy for the ``bath'' fermions $b_j$ with strength $B_j(t)$, and a term that either hops an $a_j$ to a $b_j$ fermion, or pairs the two species of fermions. 
The fermionic form \eqref{eq:HG fermionic} explains why our protocol can cool the topological and the trivial phases with similar efficiency: in both phases, a single excited fermion in the bulk can be removed by hopping into the bath.

The Hamiltonian $H_{\rm GTFI} + H_{\rm c}$ is {\it quadratic} in $a_j$ and $b_j$, which allows us to simulate the cooling process for very large system sizes in Fig.~\ref{fig:numerics} (see \cite{SM} for details of how the simulation is performed). Furthermore, the quadratic form of \eqref{eq:HG fermionic} implies that in the absence noise and in the adiabatic limit, the cooling protocol brings the system from an arbitrary initial state to the ground state of $H_{\rm GTFI}$ within one cycle \cite{SM}. This is because the cooling occurs independently for each fermionic mode at every quasi-momentum. Importantly, the protocol can also cool systems with quartic interactions between the fermions parametrically more efficiently than other known methods such as adiabatic preparation \cite{SM}. In that case, while multiple cycles are needed in order to reach the steady state, the performance of the protocol remains parametrically the same, i.e., the ground state is approached exponentially in $n_{\rm cyc}$ and $T$.

In addition, one may expect that cooling using a simulated fermionic bath to perform efficiently whenever the elementary excitations are fermionic in nature. 
The difference between gauged cooling and ordinary, ``bosonic'' cooling is most pronounced in $d>1$. 
In this case, the mapping can be done either by a Jordan-Wigner transformation, or by other methods that preserve the locality of the Hamiltonian~\cite{Bravyi_Kitaev_2002,kitaev2006anyons,Verstraete_Cirac_2005}.
To demonstrate the advantage of gauged cooling in $d>1$ fermionic systems, we analyze the cooling rate of low energy excitations in a band insulator model on a narrow 2D strip \cite{SM}. 
 We find that the rate of gauged cooling remains independent of system size while the rate of simple bosonic cooling falls off exponentially as a function of the width of the strip.
The same gauged cooling protocol could also efficiently prepare fermionic topological phases, such as a chiral two-dimensional superconductor.

To summarize, we have shown how encoding the degrees of freedom of a  system non-locally into those of a quantum simulator can accelerate the preparation of low energy states.  
The advantage, compared to using a local encoding, is dramatic in cases where the system's excitations are spatially non-local. 
In the quantum Ising model, coupling the Ising spins to a $Z_2$ gauge field allows to remove the topological domain wall excitations of the ferromagnetic phase as easily as the spin flip excitations of the paramagnetic phase. 
During the protocol, entropy is pumped to the gauge degrees of freedom, where it is removed by the measurement. The measurement outcomes are also used 
to determine the gauge transformation that relates observables of the modified (gauged) Hamiltonian to those of the original one. 

The non-local encoding of the system's degrees of freedom implies that local errors in the physical qubits, as well as measurement errors, can create highly non-local excitations in the system. However, since the non-local excitations can also be removed by our protocol, their density in the steady state is small in proportion to the error rate. Similarly, local gauge-invariant observables are not strongly affected by noise and measurement errors (see~\cite{SM} for further discussion).


The operation of our protocol becomes particularly transparent upon representing the system and auxiliary (gauge) spins in terms of fermions. The gauge degrees of freedom play the role of auxiliary fermionic sites, which are coupled to the system's sites via single fermion hopping or pairing. These auxiliary sites serve as a fermionic bath. In the fermionic language, our protocol can prepare the trivial and the topological phases of a one-dimensional superconductor with the same effort.

Cooling in our protocol is possible thanks to the non-unitary operations performed during the circuit, namely measurements or resets. 
In that respect, our protocol is reminiscent of setups that use measurements and classical feedback to prepare topological phases of matter~\cite{Kitaev_2003,Piroli2021,Verresen2021,tantivasadakarn2022hierarchy}. However, our protocol does not require knowledge of the target state; thus, the protocol works even away from ``fixed point'' Hamiltonians with zero correlation length.

On an ideal quantum simulator (without noise), the energy density within our protocol converges to the ground state value exponentially in $T$ and $n_{cyc}$. 
Thus, $T$ and $n_{cyc}$ scale logarithmically with the inverse of the required accuracy of the energy density. 
Using a circuit of fixed depth, the accuracy in the energy density is {\it independent} of the system size. 
Crucially, our protocol only relies on the presence of a gap in the spectrum of the 
{\it target} Hamiltonian~\footnote{See Ref. \cite{mathhies2022adibatic_demag} for a discussion of this fact in the non-gauged cooling case. Similar arguments apply for gauged cooling.}. 
In contrast, adiabatic preparation protocols require a finite gap along the entire adiabatic path connecting the initial (product) state to the target state. 
As a result, the error in the energy density scales as a power of the system size if a continuous phase transition has to be crossed along the way (as in the case of a topologically non-trivial target state). 

On real quantum hardware, one should optimize the number of Trotter steps. Errors due to environmental decoherence and gate infidelity increase with the number of gates performed per cycle (and hence with the number of Trotter steps), whereas adiabatic and Trotterization errors decrease with the number of steps.

Our method can be applied 
to any one-dimensional gapped fermionic system, requiring no prior knowledge of the nature of its ground state. We believe that the framework of gauged cooling can be extended to other situations of interest. For instance, it can be readily applied to the TFIM in higher dimensions, where the domain walls are $(d-1)$-dimensional objects. The same technique of introducing auxiliary gauge fields can serve to eliminate the domain walls efficiently. In $d>1$, not all configurations of the gauge field are equivalent, since configurations with different gauge fluxes are not related by a gauge transformation. However, since the ground state of the gauged model is free of fluxes, the fluxes will tend to be removed once the protocol converges to the ground state.



We also expect our algorithm to perform well in preparing symmetry-protected topological states, even if the protecting symmetry is maintained throughout the cycle. For example, this may allow efficient preparation of the AKLT state where recent work has proposed to use post-selection to reduce the gate depth \cite{Smith2023AKLT}. Finally, an important open question is whether non-local encoding and fermionic bath engineering offer any advantage over more traditional preparation methods
in critical (gapless) quantum states.



\acknowledgements

We thank Ehud Altman, Debanjan Chowdhury, Elliot Kapit, Yuri Lensky, Vadim Oganesyan, Anne Matthies, Gil Refael, Shengqi Sang, Steve Simon, and Ashvin Vishwanath for useful discussions. E.B. and A.R. were supported by CRC 183 of the Deutsche Forschungsgemeinschaft (project number 277101999, subproject A01). 
 M.R. acknowledges the Brown Investigator Award, a program of
the Brown Science Foundation, the University of Washington College of Arts and Sciences, and the Kenneth K.
Young Memorial Professorship for support.

\bibliography{References}

\end{document}


\title{Supplementary Material for: \\ 
Gauged cooling of topological excitations and emergent fermions on quantum simulators}
    
    
\author{Gilad Kishony}
\affiliation{Department of Condensed Matter Physics,
Weizmann Institute of Science,
Rehovot 76100, Israel}
\author{Mark S. Rudner}
\affiliation{Department of Physics, University of Washington, Seattle, WA 98195-1560, USA}
\author{Achim Rosch}
\affiliation{Institute for Theoretical Physics, University of Cologne, 50937 Cologne, Germany}
\author{Erez Berg}
\affiliation{Department of Condensed Matter Physics,
Weizmann Institute of Science,
Rehovot 76100, Israel}

\maketitle


\section{Cooling the TFIM with periodic boundary conditions}
\label{app: periodic}

When cooling the 1D GTFI model Eq.~(2) of the main text with periodic boundary conditions, the total $Z_2$ parity, $U=\prod_j\sigma_j^z\tau_j^z$ is conserved by the Hamiltonian $H(t) = H_{\rm GTFI} + H_{\rm c}$. However it can be affected by noise such as bit flips acting on $\sigma$ or $\tau$ degrees of freedom. 

In the absence of noise, we expect the cooling protocol to work as usual with the exception that the ground state 
will be achieved only within the fixed parity sector rather than reaching the global ground state. In the ferromagnetic phase, the ground state is found in the flux-free sector, $F = \prod_j\tau_j^z=1$ (where all domain walls can be eliminated) for both values of the total parity. In contrast, in the paramagnetic phase the ground state has $\prod_j\sigma_j^z=1$ for both values of the total parity, such that the flux $\prod_j\tau_j^z$ of the ground state has the same sign as the total parity. In both the paramagnetic and ferromagnetic phases, the difference between the ground state energies in the two parity sectors is exponentially small in the system size. In the absence of noise, if the initial state is chosen to have total parity $U=1$, the cooling protocol succeeds to find the global ground state in both phases. It is noteworthy that in the ferromagnet, the algorithm finds a Greenberger–Horne–Zeilinger state \cite{Bouwmeester1999}, which is a macroscopic superposition of the two symmetry broken states with opposite values of the order parameter $\langle \sigma^x_j\rangle$.

If noise is present, the total parity fluctuates between the two sectors. Correspondingly, in the ferromagnetic phase, the system parity $\prod_j\sigma_j^z$ fluctuates in the steady state while the flux $F$ tends to $1$ if cooling is done rapidly compared to the rate of noise. In the paramagnetic phase, the system parity $\prod_j\sigma_j^z$ tends to $1$ while the flux $F$ fluctuates between the two sectors with equal probabilities in a large system.

\begin{figure}
\begin{centering}
\includegraphics[width=0.9\columnwidth]{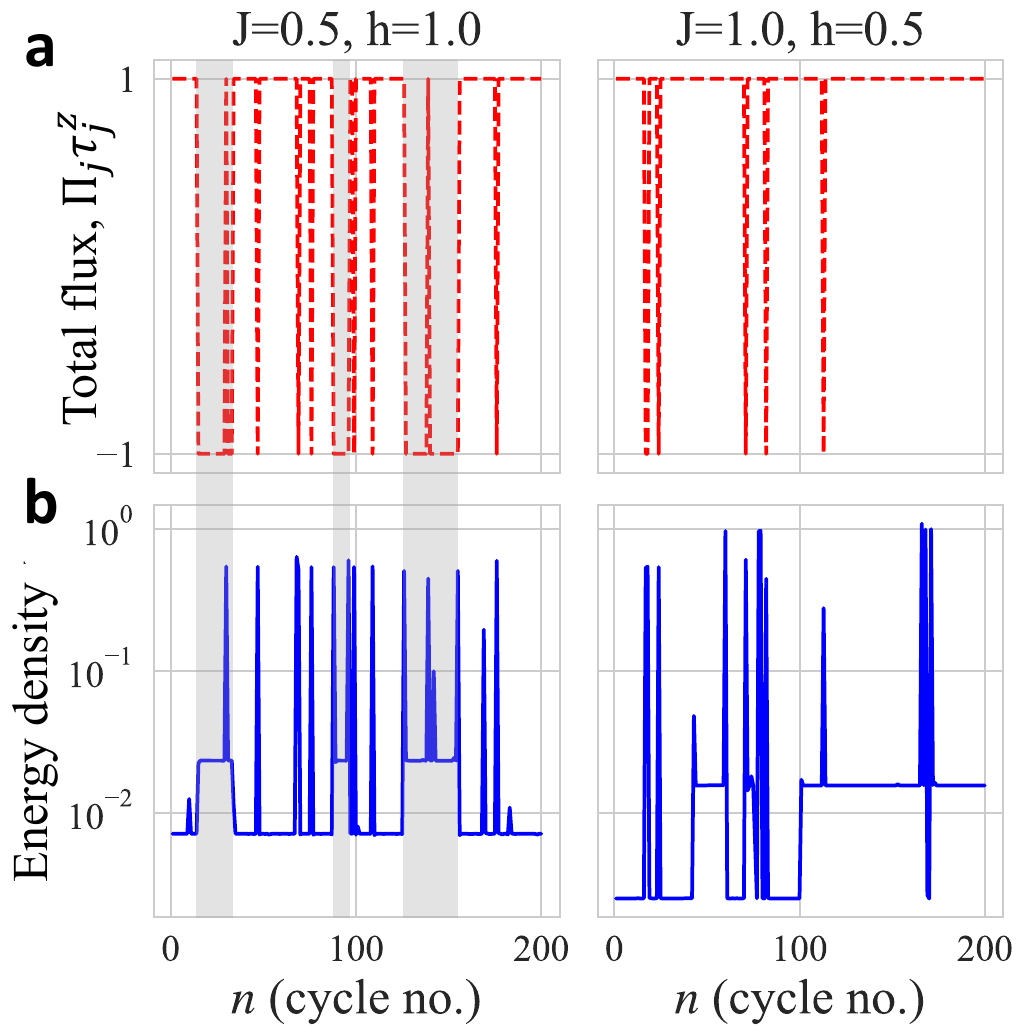}
\par\end{centering}
\caption{Cooling the 1D TFIM with a noise rate of $0.02$ errors per cycle per spin with $N=4$ sites and periodic boundary conditions in the paramagnetic phase - $J=0.5, h=1.0, V=0.0$ (left) and in the ferromagnetic phase - $J=1.0, h=0.5, V=0.0$ (right). Protocol parameters are set to $g_1=0.5$, $B_0=3$, $T=50$ and $N_t=100$ Trotter steps are performed per cycle. The measured flux $F=\prod_j\tau^z_j$ \textbf{(a)} and the energy density \textbf{(b)} are shown as a function of the cycle number. In the paramagnetic phase, the measured flux approximately coincides with the sector of total $Z_2$ parity, $U=\prod_j\sigma_j^z\tau_j^z$, which is conserved by $H(t)$ and can be flipped by noise as indicated by the gray shaded regions. In the ferromagnetic phase, the steady state flux is fixed, even when the sector of the total parity is flipped by noise.}
\label{fig: single time trace periodic}
\end{figure}

This behavior is demonstrated numerically for both phases in Fig.~\ref{fig: single time trace periodic}, in which a system with $N=4$ sites is cooled in the presence of noise. The measured flux $F$ at the end of each cycle along with the energy density are shown. Stochastic application of noise 
results in sharp 
peaks in the energy which are cooled by subsequent cycles. However, when the noise flips the total parity, further cooling reaches a steady state which is near the ground state of the new parity sector. The measured flux 
exhibits sharp changes coinciding with some of the events of stochastic noise in both phases. However, remarkably, in the paramagnetic phase of the model, the steady state value of this flux coincides with the sector of the total parity and can therefore allow one to indirectly monitor this value. On the other hand, in the ferromagnet, the flux tends to settle back at $1$ in both sectors. By 
conditioning the protocol to terminate only on 
cycles with flux $F=1$, one can obtain an approximation for the ground state of the quantum Ising model with periodic boundary conditions.

\section{Avoiding mid-circuit measurements}
\label{app: mid circuit measurements}

On some present day quantum devices, mid-circuit measurement can be slow and/or of low fidelity, and adjusting the circuit following the measurement by classical feedback is impractical. In our protocol, 
such limitations can be avoided by substituting mid-circuit measurements by resets (or fresh ancillae), if an extra ancilla qubit is added per site. This is possible since the feedback required is spatially local.

\begin{figure}
\begin{centering}
\includegraphics[trim={0cm 2cm 0cm 0cm},clip,width=(\textwidth-\columnsep)/3]{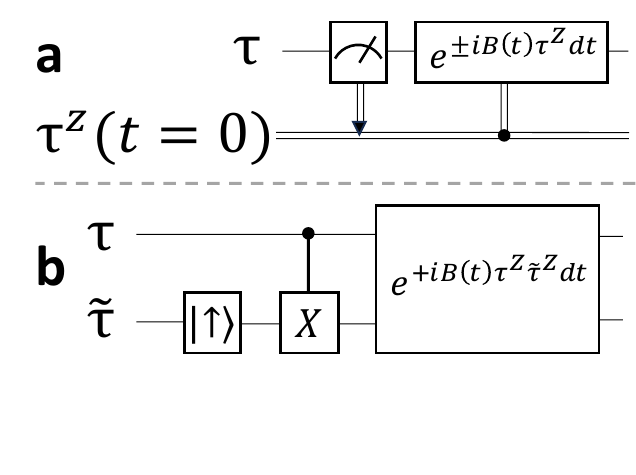}
\par\end{centering}
\caption{Substituting mid-circuit measurements by resets (or fresh ancillae). \textbf{(a)} The bath spin is measured and the sign of the coupling $B$ is adjusted accordingly. \textbf{(b)} Instead, the value of the bath spin in the $z$ basis is ``copied'' to another freshly reset qubit. The term $B$ is adjusted to include an interaction with the extra ancilla.}
\label{fig. replacing measurements with resets}
\end{figure}

In the protocol described in the main text, each cooling cycle ends with a measurement of the bath spins in the $\tau^z_j$ basis. The sign of the Zeeman field on each site is adjusted accordingly in the subsequent cycle $B_{j}(t) = B(t)\tau_{j}^{z}(t_n)$. This is illustrated in Fig.~\ref{fig. replacing measurements with resets}(a).

In the measurement-free alternative to this protocol, an ancilla spin $\tilde{\tau}_j$ is added to each site. 
Instead of the measurement of $\tau^z_j$ at the end of each cycle, we carry out the following operations:
\begin{enumerate}
\item
The auxiliary qubits are reset or freshly prepared in the state $\vert\tilde\tau_j^z=1\rangle$.
\item
A controlled-X gate with $\tau_j$ as its control and $\tilde\tau_j$ as its target is applied: $U_{\tau \tilde{\tau}} = \exp[\frac{i\pi}{4}(1-\tau^z_j)(1-\tilde{\tau}^x_j)]$. 
This essentially ``copies'' the value of $\tau_j^z$ to $\tilde{\tau}^z_j$. 
\item
The Zeeman field term in $H_{\rm c}(t)$ is replaced by a two body term acting on both $\tau_j$ and $\tilde\tau_j$, $-B_{j}(t)\tau_{j}^{z}\rightarrow-B(t)\tau_{j}^{z}\tilde\tau_{j}^{z}$.
\end{enumerate}

The circuit corresponding to this measurement-free protocol is shown in Fig.~\ref{fig. replacing measurements with resets}(b). Effectively, instead of altering the signs of the Zeeman field by classical control, they are set by quantum controlled gates conditioned on $\tilde\tau_j$. This comes at the cost of replacing the one-body spin terms $B_j$ by two body terms, although in general, the system Hamiltonian will have terms acting on more than two spins regardless.

\section{Performance in the presence of interactions}
\label{app: full state evolution with interactions}

As shown in the main text, the Hamiltonian $H(t) = H_{\rm GTFI} + H_{\rm c}(t)$ is equivalent to a free fermion Hamiltonian [Eq. (4) of the main text]. This allows for efficient classical simulations of the cooling protocol for large systems. It is important, however, to verify that our protocol is capable of cooling efficiently more generic Hamiltonians with topological excitations. 

To this end, we add to $H_{\rm GTFI}$ a term that corresponds to a quartic interaction in the fermionic representation. 
The added term conserves the parity $U$ (which is the fermion parity in the fermionic language). The simplest such term is
\begin{align}
-\sum_{j}V\sigma_{j}^{z}\sigma_{j+1}^{z}=-\sum_{j}V\left(1-2a^\dagger_j a^{\vpd}_j\right)\left(1-2a^\dagger_{j+1} a^{\vpd}_{j+1}\right).
\label{eq: interaction}
\end{align}
Adding interactions makes the system non-integrable, and we resort to numerical simulation of the stochastic Schr\"{o}dinger equation on the full Hilbert space by which we can access only small system sizes.

\begin{figure}
\begin{centering}
\includegraphics[width=0.8\columnwidth]{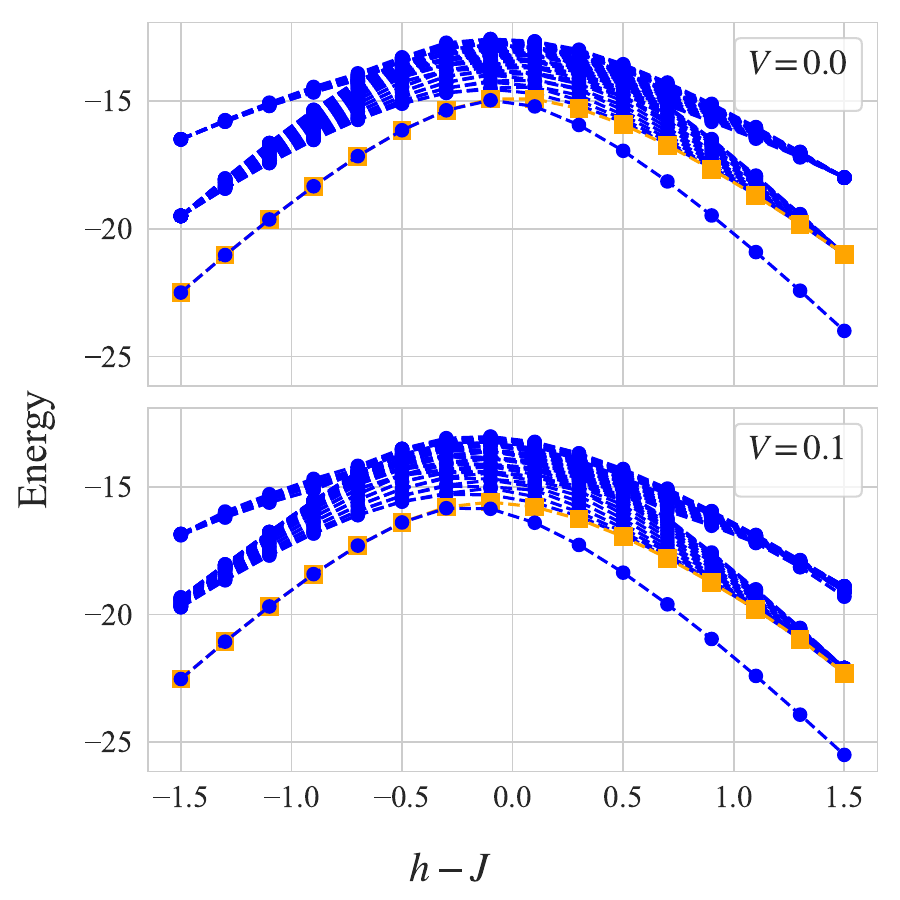}
\par\end{centering}
\caption{The spectrum of the system (decoupled from the bath) at $V=0.0$ (non-interacting fermions) and at $V=0.1$ (interacting fermions) obtained by numerical exact diagonalization of a system of size $N=16$ as a function of $h-J$ with the sum $h+J=1.5$ held fixed. The phase transition between the ferromagnet (left) and paramagnet (right) can clearly be seen as the point where the ground state degeneracy changes from two to one. The first excited state is marked by orange squares.}
\label{fig: system spectrum}
\end{figure}

We begin by mapping out the phase diagram of the model in the presence of the $V$ term, in order to locate the boundaries of the ferromagnetic and paramagnetic phases. Fig.~\ref{fig: system spectrum} shows the spectrum of $H_{\rm GTFI}$  with the added interaction of Eq.~\eqref{eq: interaction}, setting $V=0.1$, for a system with $N=16$ sites with open boundary conditions. The spectrum is shown as a function of $h-J$ along the line $h+J=1.5$. The transition from the ferromagnet to the paramagnet (or, in terms of the fermionic degrees of freedom, from the topological to the trivial phase) is identified as the point where the ground state degeneracy is lifted. We find that upon changing $V$ from $0$ to $0.1$, the transition shifts from $h-J=0$ to $h-J \approx -0.25$.


Next, we analyze the dependence of the steady state energy on the noise rate and the system size. This dependence is determined by the nature of excitations in the system. 
The dynamics of the cooling process can be understood from a simple phenomenological rate equation for the energy density during the  process~\cite{mathhies2022adibatic_demag},
\begin{equation}
    \dot{e} = A\eta - B (e/e_0)^M - C\frac{1}{N}(e/e_0),
\label{eq:rate}
\end{equation}
where $A$, $B$, $C$, $e_0$ are positive constants, $e$ is the energy density, and $N$ is the system size. The term $B$ represents a bulk cooling rate where $M$ is the number of excitations that can be removed locally in the bulk: $M=1$ for trivial excitations, and $M=2$ for Ising domain walls coupled locally to the bath. The $C$ term describes single non-local excitations which are removed near the edges of the system; this cooling rate is proportional to the number of excitations times their probability to be within a correlation length of the edge, which is proportional to $1/N$. In Eq. \eqref{eq:rate}, the discrete time evolution corresponding to the cycles of the cooling process was replaced by a continuous time evolution. 

Setting $\dot{e}=0$, Eq. \eqref{eq:rate} implies that in the thermodynamic limit, the energy of the steady state is proportional to $\eta^{1/M}$. For $M>1$, small systems with open boundary conditions should exhibit a crossover behavior: if the steady-state number of excitations in the entire system is $O(1)$ the rate of cooling through the edge dominates. In this case, the energy density of the steady state is expected to scale as $\eta N$. In contrast, in a system with local excitations ($M=1$), the energy density should be independent of $N$. 
Such behavior was indeed observed in Ref. \cite{mathhies2022adibatic_demag}: 
the steady state energy density scales as $N$ in the ferromagnet, and is approximately independent of $N$ in the paramagnet. 

\begin{figure}
\begin{centering}
\includegraphics[width=(\textwidth-\columnsep)/2]{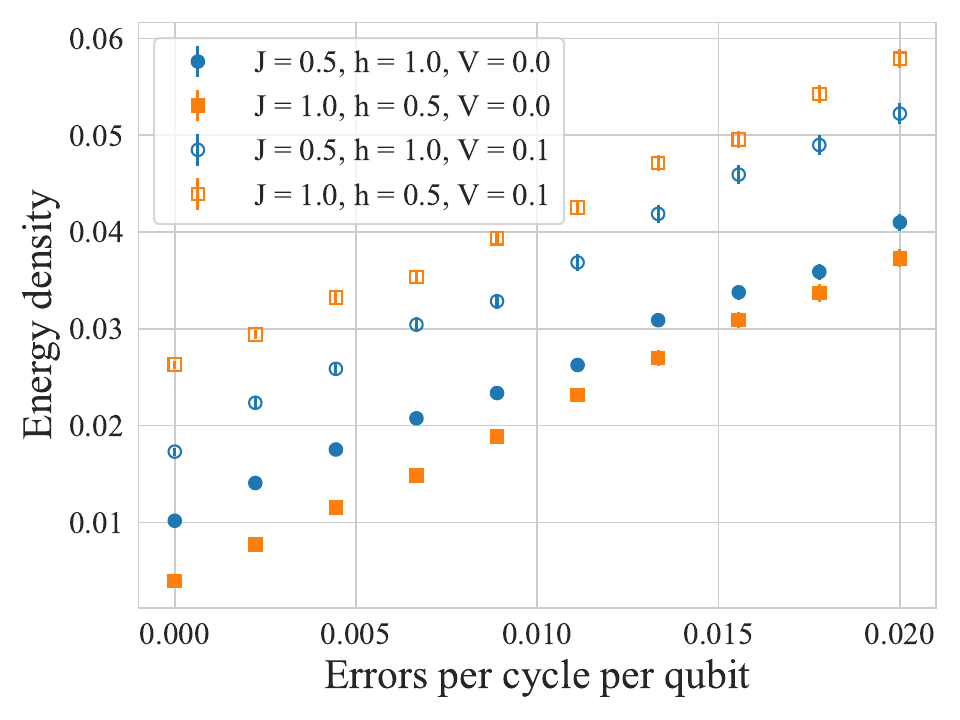}
\par\end{centering}
\caption{Cooling the 1D TFIM with $N=8$ spins in the paramagnetic phase (blue circles) and in the ferromagnetic phase (orange squares) in the presence of decoherence. Full symbols correspond to the non-interacting model at $V=0$ while empty markers are with $V=0.1$. The average system energy density measured at the end of cooling cycles is plotted as a function of the error rate. This is calculated using $N_t=100$ Trotter steps per cycle and the energies are averaged over $10,000$ cycles per data point. Protocol parameters are set to $g_1=0.5$, $B_0=3$, $T=50$. In all cases, the energy density is linear in the error rate, demonstrating that cooling is done at a rate proportional to the density of excitations.}
\label{fig. energy density vs. error rate stochastic schrodinger 8 sites V=0}
\end{figure}

Fig.~\ref{fig. energy density vs. error rate stochastic schrodinger 8 sites V=0} presents the average energy density in the steady state for our gauged cooling protocol as a function of the decoherence error rate in both phases, with $V=0$ and $V=0.1$. As in the main text, the error rate is defined as the average number of depolarizing errors per spin ($\sigma$ or $\tau$) per cycle. (For a detailed description of the depolarizing noise model, see App. \ref{sec:noise} below.) These simulations were performed for a system of size $N=8$, and data points are averaged over $10,000$ cycles. In all cases the energy density depends linearly on the noise rate, as for the non-interacting system. 

The steady state energy density is non-zero even in the absence of noise, due to adiabatic and Trotter errors. Interestingly, in both the ferromagnetic and paramagnetic phases, the energy density increases upon introducing interactions. There is no obvious correlation between the energy density and the gap between the ground state and the first excited state (see Fig.~\ref{fig: system spectrum}). It thus seems that the interacting system is more susceptible to Trotter and adiabatic errors; we leave a detailed study of these effects to a future study.

\begin{figure}
\begin{centering}
\includegraphics[width=(\textwidth-\columnsep)/2]{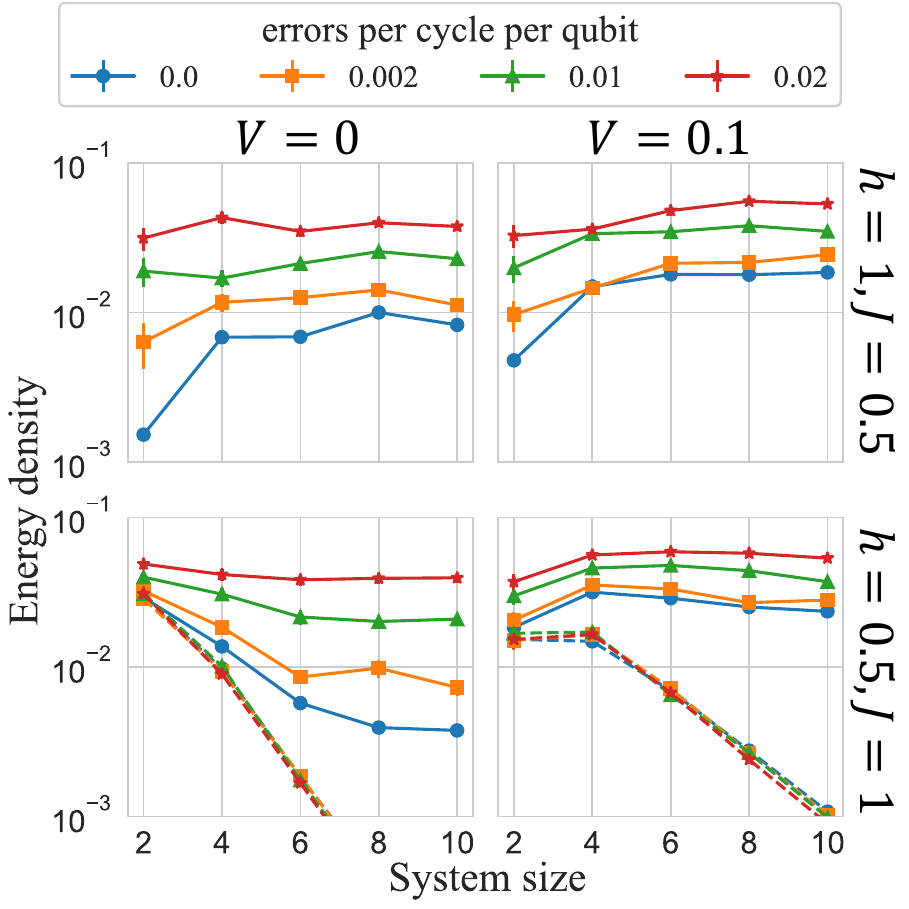}
\par\end{centering}
\caption{Finite size scaling of the steady state energy density reached by cooling the 1D TFIM in the presence of different decoherence rates. The model is taken in the paramagnetic phase $h=1.0,J=0.5$ (top) and in the ferromagnetic phase $h=0.5,J=1.0$ (bottom), and the interaction strength is set to $V=0$ (left) and $V=0.1$ (right). This is calculated using $N_t=100$ Trotter steps per cycle and the energies are averaged over $1,000$ cycles per data point. Protocol parameters are set to $g_1=0.5$, $B_0=3$, $T=50$. The energy density saturates at a value independent of system size as expected. In the ferromagnetic phase, energy density decreases with system size at small sizes due to the excitation of the slightly split Majorana edge states. Dashed lines show the average energy reached in this lowest excited state alone.}
\label{fig: finite size scaling}
\end{figure}

Fig.~\ref{fig: finite size scaling} shows the average energy density as a function of system size for different error rates in the gauged cooling protocol, with interaction strengths $V=0$ and $V=0.1$. Evidently, in all tested regimes of the system parameters, the energy density saturates at a value independent of system size, as expected if the excitations are removed locally in the bulk. This indicates that even in the presence of interactions, the gauged cooling protocol is capable of removing both single local  excitations (spin flips) and single non-local excitations (domain walls). 

An interesting feature of the ferromagnetic phase (the topological phase in the fermionic language) is the initial decrease of energy density with increasing system size for small systems, especially in the smaller noise rates. It appears that this derives from the change in the excitation energy associated with the Majorana edge states, which decreases exponentially with increasing system size. Indeed, these low energy modes are more susceptible to adiabatic errors than the bulk excitations, since their energy splitting is smaller than the bulk gap. As the system size increases, the excitation energy associated with the edge states decreases rapidly, and their contribution to the steady state excitation energy decreases. 
This explanation is supported by examining the part of excitation energy associated with the first excited state only (which corresponds to the excitation of the edge states). As expected, this contribution to the energy density decreases rapidly with system size for all noise levels.

\begin{figure}[t]
\begin{centering}
\includegraphics[width=0.8\columnwidth]{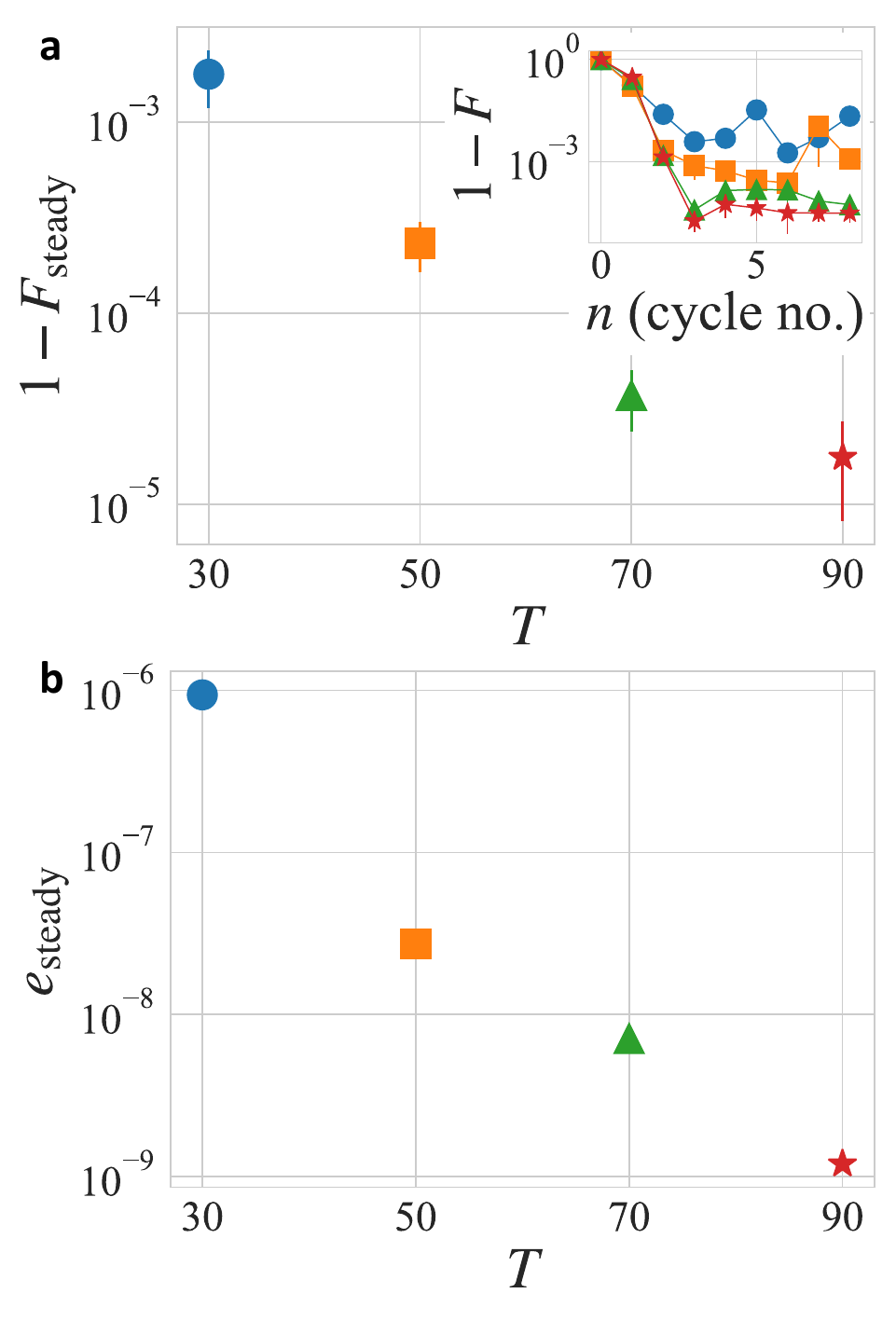}
\par\end{centering}
\caption{Cooling a system with $N=8$ sites and open boundary conditions in the ferromagnetic phase with interactions ($J=1$, $h=0.5$, $V=0.1$). The protocol parameters used were $g_1=0.5$, $B_0=3$. The initial state of the $\sigma$ spins is a random $\sigma^x$ eigenstate with even parity. The number of Trotter steps $N_t=10^5$ is sufficiently large such that the deviation of the steady state from the ground state is due to adiabatic errors and the finite size splitting of the doubly degenerate ground state $E_1-E_0\approx 0.0063$. \textbf{(a)} The steady state infidelity with the doubly degenerate ground state manifold vs. the cycle duration $T$. $100$ shots with random initialization and measurement outcomes are performed per data point. Inset: convergence to the steady state value with protocol cycles. \textbf{(b)} Energy density of the steady state vs. $T$ after postselection on even bath parity - corresponding to the sector of the system ground state.}
\label{fig:convergence with interactions}
\end{figure}

Finally Fig.~\ref{fig:convergence with interactions} demonstrates that even in the presence of interactions, cooling converges exponentially to the ground state with increasing $T$ and cycles in the absence of noise.

\section{Reaching the ground state of the non-interacting model in a single cycle}
\label{app: single cycle}

The Hamiltonian $H(t) = H_{\rm GTFI}+H_{\rm c}(t)$ given by Eq.~(4) of the main text is quadratic in the fermionic degrees of freedom. Moreover, in the first cooling cycle the $\tau$ spins are initialized in the fully polarized state, such that the Zeeman field has uniform signs and the Hamiltonian is translationally invariant. We find that if the time evolution is perfectly adiabatic, our protocol is capable of bringing the system to the ground state after a single cycle, independently of its initial
state. To see this, we consider the Hamiltonian $H(t)$ in momentum space: 
\begin{equation}
H(t)=2\sum_{k>0}C_{k}^{\dagger}H^{\vpd}_{k}(t)C^{\vpd}_{k},
\end{equation}
where $C_{k}^{\dagger}=\left(a_{k}^{\dagger},a_{-k}^{\vpd},b_{k}^{\dagger},b_{-k}^{\vpd}\right)$, $a_k,b_k$ are the Fourier transforms of $a_j,b_j$ respectively, and $H_{k}(t)$ has the form
\begin{equation}
H_{k}(t)=\left(\begin{array}{cc}
H_{k}^{\rm{sys}} & G_k(t)\\
G_k^\dagger(t) & H^{\rm{bath}}(t)
\end{array}\right).\label{eq:Hk}
\end{equation}
Note that the upper left $2 \times 2$ block $H^{\rm{sys}}_k$ acts on the {\it system} degrees of freedom ($\{a_j\}$), while the lower right block $H^{\rm{bath}}(t)$ acts on the fermionic {\it bath} degrees of freedom ($\{b_j\}$); the off-diagonal blocks $G_k(t)$ represent the system-bath coupling.

The four instantenous eigenvalues of $H_{k}(t)$,
sorted in ascending order, are denoted $E_{n=1,\dots,4,k}(t)$. Crucially, the gaps between the energies of the lowest and highest states, $E_{1,k}(t)$ and $E_{4,k}(t)$, and the rest of the spectrum do not
close for any value of $k$ and $t$. We denote the fermionic creation operator for the eigenmode
corresponding to $E_{n,k}(t)$ by $f_{n,k}^{\dagger}(t)$.

In the beginning of the cooling protocol, the coupling $g(t=0)$ vanishes,
and the bath is initialized such that $\langle b_{j}^{\dagger}b_{j}^{\vpd}\rangle=0$
for all $j$. In momentum space, this corresponds to $\langle b_{k}^{\dagger}b_{k}^{\vpd}\rangle=0$.
Initially, when $B(t)$ is the largest energy scale in the system and $g(t) = 0$, the lowest and highest energy eigenmodes of the system arise from the eigenstates of the (decoupled) block $H^{\rm{bath}}$: $f_{1,k}(t=0)=b_{-k}^\dagger$
and $f_{4,k}(t=0)=b_k$. The initial conditions above correspond to $\langle f_{1,k}^{\dagger}f_{1,k}^{\vpd}\rangle=1$, $\langle f_{4,k}^{\dagger}f_{4,k}^{\vpd}\rangle=0$,
whereas $\langle f_{n,k}^{\dagger}f_{m,k}^{\vpd}\rangle=0$ for $n=1,4$ and
$m=2,3$. 
In contrast,
the middle block of the single-particle density matrix, $\langle f_{n,k}^{\dagger}f_{m,k}^{\vpd}\rangle$
with $n,m=2,3$, depends on the initial state of the system. 

Under perfectly adiabatic evolution, the diagonal elements of the single-particle density matrix,
$\langle f_{n,k}^{\dagger}(t)f_{n,k}^{\vpd}(t)\rangle$, are preserved.
At the end of the cycle (when $g(t)$ and $B(t)$ are ramped to zero), $f_{1,k}^{\dagger}(T)$ and $f_{4,k}^{\dagger}(T)$
are the eigenmodes of $H_k^{\rm{sys}}$.
As noted above, $f_{1,k}(T)$ and $f_{4,k}(T)$ thus correspond to the ground and excited bands of the system.
Adiabaticity of the evolution together with the initial condition $\langle f_{1,k}^{\dagger}f_{1,k}^{\vpd}\rangle=1$ gives $\langle f_{1,k}^{\dagger}(T)f_{1,k}^{\vpd}(T)\rangle=1$ and $\langle f_{4,k}^{\dagger}(T)f_{4,k}^{\vpd}(T)\rangle=0$.
Hence, the system is transferred to its ground state at the end of the cycle, regardless of its initial state.
At the same time, the (arbitrary) initial single-particle
density matrix of the system is ``copied'' into the density matrix
of the bath fermions [which now correspond to $f_{2,k}(T)$ and $f_{3,k}(T)$].
Thus, in the perfectly adiabatic limit, the system reaches its ground
state after a single cycle. 

Remarkably, the ground state is reached after a single cycle both in the trivial and the topological phase of the one-dimensional superconductor Hamiltonian given by Eq.~(4) of the main text. Hence, our protocol is capable of preparing a topological phase of fermions using a finite depth circuit. This can be viewed as follows: the system and the bath fermions are initially in a trivial (product) state. After the time evolution, the system and the bath are both in the topological phase, such that overall, the system remains trivial. Hence, no gap needs to close during the unitary evolution. However, at the end of the cycle, the system and the bath are {\it decoupled} from each other. Hence, the system has been prepared in the topological state. The possibility of preparing topological phases in this fashion has been pointed out in Ref.~\cite{tantivasadakarn2022hierarchy}; our protocol offers an explicit, generic way to achieve this.

\section{Cooling 
free fermions on a 
2D strip}

Beyond the advantage gauged cooling offers in the ferromagnetic phase of the 1D TFIM, we expect that this approach also offers a parametric advantage in the preparation of the ground state of any system of fermions in $d>1$, 
compared with 
straightforward bosonic cooling. The reason is that the gauged cooling protocol can remove a single elementary fermionic excitation, despite the fact that it is represented as a non-local object in terms of the physical (bosonic) qubits of the quantum simulator. In contrast, using an ordinary bosonic bath can only remove a pair of fermionic excitations. 

\begin{figure}
\begin{centering}
\includegraphics[width=0.9\columnwidth]{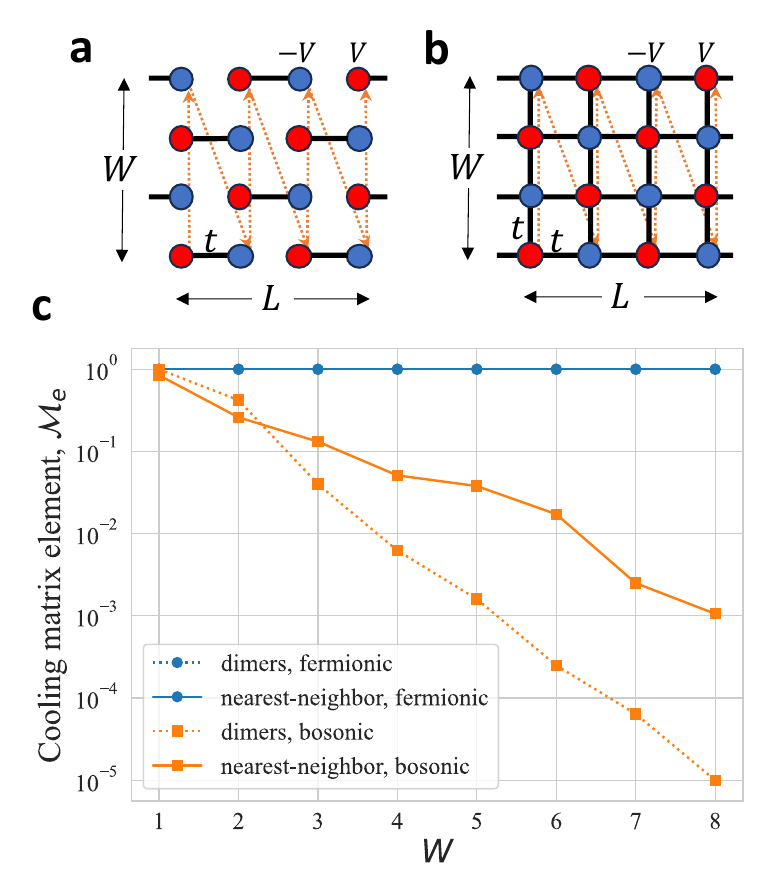}
\par\end{centering}
\caption{
Cooling topologically trivial free fermions on a narrow 2D strip. \textbf{(a)} A simple model of horizontally dimerized fermions given by Eq.~\eqref{eq: dimerized model}. \textbf{(b)} A model of uniform nearest-neighbor hopping on a square lattice given by Eq.~\eqref{eq: square lattice model}. The sites are enumerated for the Jordan-Wigner transformation as illustrated in orange. A staggered onsite potential $V$ opens a gap. \textbf{(c)} The sum of matrix elements controlling the perturbative cooling rate from the first excited state, Eq.~\eqref{eq: perturbative cooling rate}, vs. the width of the strip $W$ for gauged (fermionic) and simple (bosonic) cooling of each of the models. Parameters $L=20$, $V=0.5$, $t=1$ are held fixed, and the boundary conditions are periodic in $x$ and open in $y$ as drawn.}
\label{fig: cooling 2d strip}
\end{figure}

To demonstrate this, we consider two gapped models of free fermions (band insulators) in two dimensions: horizontally dimerized fermions as shown in Fig.~\ref{fig: cooling 2d strip}(a), and fermions with uniform nearest-neighbor hopping on a square lattice as shown in Fig.~\ref{fig: cooling 2d strip}(b). In both models, fermionization is done by a Jordan-Wigner transformation with the sites enumerated in a zigzag pattern as illustrated. A staggered onsite potential $V$ is included to open a gap. 
We consider a system defined on an $L\times W$ strip with periodic boundary conditions in the $x$ direction and open boundary conditions in the $y$ direction.
For the dimerized model, the Hamiltonian is given by
\begin{equation}
H_{\rm a} = -\sum_{i=1}^L \sum_{j=1}^W \left[ta^\dagger_{i,j}a^{\vpd}_{i+(-1)^{i+j},j}+(-1)^{i+j}Va^\dagger_{i,j}a^{\vpd}_{i,j}\right],
\label{eq: dimerized model}
\end{equation}
while the square lattice model is given as

\begin{align}
H_{\rm b} = &-\sum_{i=1}^L \sum_{j=1}^W \left[ta^\dagger_{i,j}a^{\vpd}_{i+1,j}+ta^\dagger_{i,j}a^{\vpd}_{i,j+1}+h.c.\right]\nonumber\\
&-\sum_{i=1}^L \sum_{j=1}^W (-1)^{i+j}Va^\dagger_{i,j}a^{\vpd}_{i,j},
\label{eq: square lattice model}
\end{align}
where $t$ is the hopping strength.

We compare the efficiency of preparing the ground states of $H_{\rm a,b}$ by either gauged (fermionic) cooling orbosonic cooling (of the type described in \cite{mathhies2022adibatic_demag}) in the perturbative limit, assuming a small
system-bath coupling $g$. 
In the bosonic case, each one of the system's qubits $\sigma_{i,j}$ (related to the fermionic operators $a_{i,j}$ by a Jordan-Wigner transformation) is coupled to a bath qubit $\tau_{i,j}$ by a local coupling of the form $g \sigma_{i,j}^x \tau^x_{i,j}$. In the gauged cooling protocol, in contrast, the Jordan-Wigner transformation includes the bath sites, such that after fermionization the bath sites are also represented by fermions, $b_{i,j}$, and the system-bath coupling corresponds to a coherent fermionic hopping of strength $g$ between the system and the bath [analogously to Eq. (4) in the main text]. Note that in either case, the system-bath coupling does not conserve the number of fermions in the system, such that the $U(1)$ charge conservation of the target Hamlitonian does not impose constraints on the dynamics during the cooling process.

As described in \cite{mathhies2022adibatic_demag}, in the perturbative limit, the approach to the steady state is controlled by 
the rate at which low energy excitations are removed through the system's interaction with the bath. Using Fermi's golden rule, this rate is proportional to the cooling matrix element
\begin{equation}
\mathcal{M}_e = \sum_{i,j} |\bra{0}O_{i,j}\ket{e}|^2,
\label{eq: perturbative cooling rate}
\end{equation}
where $\ket{0}$ is the ground state of the system, $\ket{e}$ is the lowest energy excited state, and $O_{i,j}$ is the operator of the system that couples to the local bath degree of freedom. 
In our case, $\vert e \rangle$ is either an added fermion at the bottom of the conduction band, or a hole at the top of the valence band. 
Without loss of generality, we focus on the former case of an added fermion excitation. 
Then, for bosonic cooling, $O^{\rm b}_{i,j}=\sigma_{i,j}^-=\prod_{k<i,l} (-1)^{a^\dagger_{k,l} a^{\vpd}_{k,l}}\prod_{l<j} (-1)^{a^\dagger_{i,l} a^{\vpd}_{i,l}}a_{i,j}$, whereas for gauged cooling, $O^{\rm f}_{i,j}=a_{i,j}$. 


Figure~\ref{fig: cooling 2d strip}(c) shows this proxy for the cooling rate vs.~the width of the strip $W$ for fermionic/bosonic cooling of each of the models. In both cases, the bosonic rate decays exponentially with $W$ while the fermionic one remains constant. This is due to the fact that when cooling with the bosonic operator $\sigma_{i,j}^-$ on each site $(i,j)$, a fermionic excitation at site $(i,j)$ can be removed, but the Jordan-Wigner string anti-commutes with $O(W)$ hopping terms and can leave behind new excitations at those locations. We have verified that the dependence of the matrix element on $W$ is similar for other higher excited states.

We can capture this behavior analytically in the dimerized model, $H_{\rm a}$, by noting that a low energy excited state corresponds to adding or removing a fermion to a single dimer. Each term in the sum in Eq.~\eqref{eq: perturbative cooling rate} therefore decomposes as a product of terms, one for each dimer. The only non-zero terms in the sum are the two terms with $(i,j)$ being one of the sites of the excited dimer. All dimers intersected by the Jordan-Wigner string zero or two times trivially give a multiplicative factor of $1$. In the latter case, this factor is evaluated as
\begin{equation}
|\bra{0_d}(-1)^{a^\dagger_{1,d} a^{\vpd}_{1,d}}(-1)^{a^\dagger_{2,d} a^{\vpd}_{2,d}}\ket{0_d}|^2=1,
\end{equation}
with $\ket{0_d}$ the ground state of the dimer $d$ and $a_{1,d}, a_{2,d}$ the fermionic operators within the dimer. This is equal to $1$ because the ground state has well defined parity on the dimer. The dependence on the width $W$ comes from the $N_W\approx W$ dimers intersected once by the Jordan-Wigner string (counting both dimers near the excitation and dimers at $i=0$). Each of these gives a multiplicative factor
\begin{equation}
|\bra{0_d}(-1)^{a^\dagger_{m,d} a^{\vpd}_{m,d}}\ket{0_d}|^2 = 
\left(\frac{t^2-\left(V+\varepsilon\right)^2}{t^2+\left(V+\varepsilon\right)^2}\right)^2<1,
\end{equation}
where $m\in\{1,2\}$ and $\varepsilon=\sqrt{t^2+V^2}$. We obtain the exponential decay of the cooling matrix element as a function of the width of the strip
\begin{equation}
\mathcal{M}_e \sim 
\left(\frac{t^2-\left(V+\varepsilon\right)^2}{t^2+\left(V+\varepsilon\right)^2}\right)^{2W}.
\end{equation}
In the dimerized case, the exponential suppression of the matrix element can be circumvented by a different choice of the Jordan-Wigner string, such that there is at most one dimer for which one site belongs to the string and the other site does not. Clearly, in the model with uniform hopping (and more generally, any model in which the Hamiltonian is not a sum over disconnected dimers), no such choice of the string is possible, and the gauged cooling is always superior. 

Finally, we note that the advantage of the gauged cooling protocol over the bosonic one is expected to persist even in the presence of interactions between the fermions, as long as the lowest-energy fermionic excitation has a finite overlap with an addition or removal of a single fermion. In the
noiseless case and for interacting models, we expect exponential cooling
efficiency in both in ordinary and topological band-insulators or
superconductors. In this case the lowest-energy excitations are fermions
which can hop directly to the simulated fermionic bath sites.


\section{Smooth evolution of time dependent couplings}
\label{app: smooth curves}

In order to approach the adiabatic limit exponentially with increasing cooling cycle duration $T$, we choose the time dependence of the couplings $B$ and $g$ to have finite derivatives at all orders. Explicitly, we define a smooth step function $\mathcal{S}(x)$ as

\begin{align}
\mathcal{S}(x) = \begin{cases}
			0, & x<0\\
            1, & x>1\\
            \frac{1}{1+e^{\frac{1}{x}+\frac{1}{x-1}}}, & \text{otherwise}
		 \end{cases}.
\end{align}

Using this, we express $B(t)$ and $g(t)$ as

\begin{align}
B(t)=B_0\left(1-\mathcal{S}(\frac{t}{T})\right),
\end{align}

\begin{align}
g(t)=g_1\mathcal{S}(\frac{t}{t_1})\left(1-\mathcal{S}(\frac{t-t_2}{T-t_2})\right).
\end{align}

These curves are shown in Fig.~2(a) of the main text for $t_1=\frac{1}{4}T$, $t_2=\frac{3}{4}T$ as chosen in our numerical simulations.

\section{Single-particle density matrix simulation of free fermions with density measurements and resets}
\label{app: free fermion single-particle density matrix}

In general, classical simulation of the dynamics of a quantum many-body system requires
memory and time resources that scale exponentially in the system size. Remarkably,
the evolution of the single-particle density matrix of a fermionic
system undergoing a combination of free (quadratic) dynamics with
local density measurements/resets can be traced with polynomial cost.
Knowledge of the evolution of single-particle density matrix allows
the calculation of any quadratic observables, and the energy in particular.
Furthermore if decoherence is treated by the stochastic application of Pauli noise on the spin degrees of freedom, these events can be mapped to unitary evolution under a quadratic fermionic Hamiltonian by introducing one additional auxiliary site to the system, thus making these processes tractable as well. This has been exploited to simulate large systems with $N=100$ unit cells in Fig.~3 of the main text.

Below, we describe the method used to carry out these simulations. 

\subsection{Free evolution}

Starting from the Hamiltonian given by Eq.~(4) of the main text, it is convenient to transform to a Majorana representation. The Majorana operators $c_{j=1,\dots,4N}$ ($N$ is the system size) are given by:
\begin{equation}
c_{4(j-1)+s}=\begin{cases}
a_{j}+a_{j}^{\dagger}, & s=1,\\
\frac{1}{i}\left(a_{j}-a_{j}^{\dagger}\right), & s=2,\\
b_{j}+b_{j}^{\dagger}, & s=3,\\
\frac{1}{i}\left(b_{j}-b_{j}^{\dagger}\right), & s=4.
\end{cases}
\end{equation}
These operators satisfy $\{c_i,c_j\}=2\delta_{i,j}$. 
The Hamiltonian in Eq.~(4) of the main text can be written in the form
\begin{align}
H=\frac{i}{4}\sum_{i,j}c_{i}M_{i,j}c_{j}=\frac{i}{4}c^{T}Mc,
\end{align}
where the matrix $M$ satisfies
\begin{align}
M^{\dagger}=-M.
\end{align}

In the Heisenberg picture, the evolution of the fermions is given
by
\begin{align}
\partial_{t}c_{i}=i\left[H,c_{i}\right]=\sum_{j}M_{i,j}c_{j}.
\end{align}
Which is solved by the time ordered exponentiation
\begin{align}
c(t)=U^{\dagger}(t)c(0),
\end{align}
\begin{align}
U^{\dagger}(t)=T\left(\exp\left(\int_{0}^{t}M(t')dt'\right)\right).
\end{align}
This can be solved numerically in polynomial time by a Suzuki-Trotter decomposition
or by any other method of solving ordinary differential equations. In our numerical simulations we choose to use second order Trotterization, making $N_t$ steps per cooling cycle.

We denote the single particle density matrix by $R_{i,j}=\left\langle ic_{i}c_{j}\right\rangle $
and find
\begin{align}
R(t)=U^{\dagger}(t)R(0)U(t).
\end{align}

\subsection{Density measurements}

Measurement of the bilinear operator $ic_i c_j$ ($i\ne j$) and averaging over the measurement outcome is described by the application of the following quantum channel to the density matrix, $\rho$: 
\begin{align}
\rho\rightarrow\rho'=\sum_{\pm}M_{i,j,\pm}\rho M_{i,j,\pm}^{\dagger},\nonumber\\
M_{i,j,\pm}=\frac{1}{2}\left(1\pm ic_{i}c_{j}\right).
\end{align}

After the measurement of $ic_ic_j$, the single particle density matrix is transformed as
\begin{align}
 & R_{l,m}\rightarrow R'_{l,m}={\rm Tr}(\rho'ic_{l}c_{m})\nonumber\\
 & ={\rm Tr}(\sum_{\pm}\frac{1}{2}\left(1\pm ic_{i}c_{j}\right)\rho\frac{1}{2}\left(1\pm ic_{i}c_{j}\right)ic_{l}c_{m}).
\end{align}
Direct evaluation of this expression for different possibilities
of coincidence of the indices $i,j,l,m$ yields
\begin{align}
R'_{l,m}&=\left(1-\delta_{l,i}-\delta_{l,j}\right)R{}_{l,m}\left(1-\delta_{m,i}-\delta_{m,j}\right)\nonumber\\
&+\left(\delta_{l,i}+\delta_{l,j}\right)R{}_{l,m}\left(\delta_{m,i}+\delta_{m,j}\right).
\label{eq:meas}
\end{align}

Importantly, Eq. \eqref{eq:meas} shows that the evolution of the single-particle density matrix $R$ under a measurement of a quadratic operator in the fermions depends only on $R$ itself, and does not require knowledge of higher-order correlation functions~\cite{Wampler2022}.


\subsection{Density resets}

Resetting $ic_{i}c_{j}\rightarrow1$ for some $i\neq j$ is equivalent to performing
a measurement of the occupation $ic_{i}c_{j}$ and in the case of
finding a $-1$ result, applying $c_{j}$ to the state to fix the
outcome. This process is described by the following Kraus operators
\begin{align}
\tilde{M}_{i,j,+}&=\frac{1}{2}\left(1+ic_{i}c_{j}\right),\nonumber\\
\tilde{M}_{i,j,-}&=c_{i}\frac{1}{2}\left(1-ic_{i}c_{j}\right)=\frac{1}{2}\left(c_{i}-ic_{j}\right).
\end{align}
In this case, we find
\begin{align}
R'_{l,m}=\left(1-\delta_{l,i}-\delta_{l,j}\right)R{}_{l,m}\left(1-\delta_{m,i}-\delta_{m,j}\right)+R_{l,m}^{0},
\end{align}
where $R_{l,m}^{0}$ is the enforced single particle density
matrix of the reset subsystem
\begin{align}
R_{l,m}^{0}=\delta_{l,i}\delta_{m,j}-\delta_{l,j}\delta_{m,i}+i\delta_{l,i}\delta_{m,i}+i\delta_{l,j}\delta_{m,j}.
\end{align}

\subsection{Simulated noise}
\label{sec:noise}
In our numerics, we model environmental decoherence as depolarizing noise. 
This is implemented within the stochastic Schr\"{o}dinger equation method~\cite{Dalibard1992,Dum1992} by applying an error 
before each Trotter step of the time evolution and for each spin ($\{\sigma_j,\tau_j\}$) with probability of $\eta/N_t$. Whenever an error is chosen to occur, a $x$, $y$, or $z$ Pauli operator is chosen at random and applied to the state. 

Applying $z$ errors, (on either $\sigma$ or $\tau$ spins) is easy as they can be written as exponentials of quadratic fermionic operators as follows:
\begin{equation}
\sigma_{j}^{z}=e^{i\pi a^\dagger_j a^{\vpd}_j},\quad
\tau_{j}^{z}=e^{i\pi b^\dagger_j b^{\vpd}_j}.
\end{equation}
Such operators map a Slater determinant to another Slater determinant, and transform a general fermion operator to a linear combination of fermion operators. Hence, within a stochastic Schr\"{o}dinger simulation, each trajectory is a time-dependent Slater determinant, and the time evolution can be calculated efficiently. 

On the other hand, $x$ and $y$ errors of the spins correspond to complicated non-local operators in terms of the fermions. 
For example, a bit flip error on a $\sigma_j$ spin amounts to the application of $\sigma^x_j$, which becomes non-local in terms of the fermions under the Jordan-Wigner transformation of Eq.~(3) of the main text. We note, however, that the product of any two such operators can be written in terms of the fermions as exponentials of quadratic operators, for instance for $i<j$

\begin{align}
\sigma_{i}^{x}\sigma_{j}^{x}=-e^{i\pi\sum_{i\leq l<j}\left(a_{l}^{\dagger}a^{\vpd}_{l}+b_{l}^{\dagger}b^{\vpd}_{l}\right)}\left(a_{i}^{\vpd}+a_{i}^{\dagger}\right)\left(a_{j}^{\vpd}+a_{j}^{\dagger}\right)\nonumber\\
=e^{i\pi\sum_{i\leq l<j}\left(a_{l}^{\dagger}a^{\vpd}_{l}+b_{l}^{\dagger}b^{\vpd}_{l}\right)}e^{-\frac{\pi}{2}\left(a_{i}^{\vpd}+a_{i}^{\dagger}\right)\left(a_{j}^{\vpd}+a_{j}^{\dagger}\right)}.
\label{eq:exponential}
\end{align}

To overcome the obstacle of applying single Pauli $x$ or $y$ errors, we extend the system by an additional fictitious spin which we identify with $\tau_0$ and correspondingly with the fermionic operators $b_0$, $c_0$, and $c_{-1}$. 
The additional site is initialized in the state $\vert \tau_0^z=1\rangle$. 
Whenever there is an $x$ or $y$ error on $\sigma_j$ or $\tau_j$, we apply the operator $\tau^x_0 \hat{O}$, where $\hat{O} = \sigma^x_j$, $\sigma^y_j$, $\tau^x_j$, or $\tau^y_j$. In terms of the Jordan-Wigner fermions, this unitary operation is an exponential of a quadratic operators as in Eq. \eqref{eq:exponential}, and can be applied easily to a Slater determinant wavefunction. 
Since the spin $\tau_0$ does not interact with the rest of the system, as long as it is initialized in a disentangled state it will remain disentangled for the duration of the simulation. This way, we can perform simulations in the presence of depolarizing noise for large systems, as in Fig.~3(b) of the main text. 

We note that, aside from random errors in the gates, there could also be errors in the measurements. In the gauged cooling protocol, the measurements of the gauge degrees of freedom $\{\tau^z_j\}$ can be imperfect. An error in the measurement of $\tau^z_j$ between the cycles results in a wrong initialization of $B_j$ in the following cycle, which leads to the creation of additional excitations. These excitations are cooled in subsequent cycles, and the resulting contribution to the energy density in the steady state is linear in the probability for errors in the measurements. 
For the purpose of estimating long-ranged fermionic expectation values of the form $\langle a^\dagger_i a_j^{\vpd} \rangle$ with $|i-j|\gg 1$ (or, equivalently, long-ranged correlations of the order parameter of the original TFIM), measurement errors of $\{\tau^z_j\}$ are more detrimental. This is because such operators involve a string of the form $\prod_{i<l<j} \tau^z_l$, which becomes strongly sensitive to measurement errors in individual $\tau^z_l$'s as the length of the string increases. In the fermionic context, this is a generic feature of encoding fermionic degrees of freedom into bosonic qubits, and not specific to our protocol. This issue does not arise for any local observable, such as the energy, or a correlation function of bosonic observables.

\bibliography{References}